\documentclass[journal]{IEEEtran}

\ifCLASSINFOpdf
\else
\fi

\IEEEoverridecommandlockouts
\usepackage{cite}
\usepackage{amsmath,amssymb,amsfonts}
\usepackage{algorithmic}
\usepackage{graphicx}
\usepackage{textcomp}
\usepackage{xcolor}
\usepackage{tabularx}
\usepackage{pgfplots}
\pgfplotsset{compat=1.5}
\usepackage{pgfplotstable}
\usepgfplotslibrary{statistics}
\pgfplotsset{grid style={dotted,gray}}
\usepackage{tikz}
\definecolor{bblue}{HTML}{4F81BD}
\usepackage{soul}

\def\BibTeX{{\rm B\kern-.05em{\sc i\kern-.025em b}\kern-.08em
    T\kern-.1667em\lower.7ex\hbox{E}\kern-.125emX}}

\begin{document}

\title{Two-Tier Multi-Rate Slotted ALOHA for OWC/RF-Based IoT Networks}

\author{Milica Petkovic,~\IEEEmembership{Member,~IEEE,}
        Dejan Vukobratovic,~\IEEEmembership{Senior Member,~IEEE,} \\
        Andrea Munari,~\IEEEmembership{Member,~IEEE,} and~Federico Clazzer,~\IEEEmembership{Member,~IEEE}
 \thanks{M.~Petkovic and D. Vukobratovic are with University of Novi Sad, Faculty of Technical Science, 21000 Novi Sad, Serbia (e-mails:  milica.petkovic@uns.ac.rs;  dejanv@uns.ac.rs).}
 \thanks{A. Munari and F. Clazzer are with Institute of Communications and Navigation of 
the German Aerospace Center (DLR), 82234 Wessling, Germany (e-mails: andrea.munari@dlr.de; federico.clazzer@dlr.de).}
}

\maketitle

\begin{abstract}
We consider a massive Internet of Things (IoT) scenario where indoor IoT devices access the network via optical wireless communication (OWC) IoT systems that relay data via a backhaul radio frequency (RF) low-power wide-area network (LP WAN). We propose a novel two-tier multi-rate Slotted ALOHA (SA) system model to design and analyse such hybrid OWC/RF networks. For a particular hybrid OWC/RF setup, we present an in-depth numerical investigation that provides insights into the proposed two-tier multi-rate hybrid OWC/RF SA system design.  
\end{abstract}


\IEEEpeerreviewmaketitle

\section{Introduction}

\IEEEPARstart{B}{eyond} 5G networks aim to provide energy and spectrally efficient massive IoT connectivity for extreme connection densities \cite{ref01}. The main challenge lies in providing wide-area coverage for sporadic and unpredictable transmission of short packets from a vast number of devices \cite{ref03}. Low-power wide area networks (LP WAN) such as LoRa and NB-IoT are recent massive IoT solutions operating in sub-GHz radio frequency (RF) bands \cite{ref04}. Due to short packets and unpredictable device activity, their system design relies on cost-effective random access (RA) protocols, such as the variants of Slotted ALOHA (SA) \cite{ref1,ref02,ref05}.

To relieve pressure from conventional RF-based LP WANs, optical wireless communications (OWC) have been recently considered as an emerging complementary technology. For indoor environment, OWC technology operating in visible or infrared (IR) spectrum is a suitable license-free short-range connectivity solution \cite{OWC_MATLAB}.
OWC IoT relies on low-cost OWC transceivers to provide low-rate indoor IoT connectivity, either in a stand-alone or a hybrid OWC/RF setup. Indoor OWC IoT based on the SA policy has been recently analyzed in \cite{ref10, ref10a, ref10b, ref11, ref12, ref13, ref14}. 

In this paper, we propose and analyse a hybrid OWC/RF massive IoT network based on a novel \emph{two-tier multi-rate} SA system design. The first tier comprises a collection of isolated indoor OWC IoT networks connected to a network infrastructure via the second-tier outdoor RF-based LP WAN. OWC IoT devices contend for access using SA in their respective indoor OWC cells and, if successful, their packets are relayed to the second tier where they contend for SA-based access via the outdoor LP WAN (Fig. \ref{model}). The SA protocol at the LP WAN tier operates at $M$ times higher slot rate relative to the OWC tier, where judicious adaptive selection of factor $M$ based on the operating parameters of the hybrid OWC/RF IoT system is crucial for optimized system performance.  




The contribution of the paper is summarized as follows:

1) We consider a novel architecture for future massive IoT which combines short-range OWC and LP WAN technologies. The motivation is to reduce the connection density of RF-based LP WANs by offloading indoor devices to the OWC-based IoT. 

2) The proposed architecture is based on two-tier SA protocols, previously considered in the context of sensors networks\cite{Zheng_2007}. Due to OWC-RF data rate difference, we expand this model into two-tier \emph{multi-rate} SA applicable across the number of emerging use cases (e.g., drones or low-Earth orbit satellites that connect remote IoT devices to the network).

3) We present detailed analysis of packet error/outage rates at both OWC and LP WAN tiers. Combining these results into the overall system throughput, we use numerical results to provide guidelines for the system design. We emphasize the benefit of the \emph{multi-rate} SA design, where adaptive control of the slot rate factor $M$ notably improves the system throughput. We consider the optimal indoor OWC cell design by exploring the effects of OWC cell parameters on the system performance.

4) For the proposed two-tier multi-rate SA system, we consider concrete OWC and RF models. Specifically, we assume SA with capture and multi-packet reception (MPR) for indoor OWC tier, and LoRa-inspired model proposed in \cite{loraref} for outdoor LP WAN tier. Note that the model is sufficiently general to include other specific models at both tiers.

In our previous work \cite{ref12}, we focused on the design of  single indoor OWC IoT cell. In a parallel work \cite{ref13},  we considered two-tier OWC/RF massive IoT setup and analysed its fundamental behaviour using the tools from stochastic geometry, while in \cite{ref14}  two-tier OWC/RF system is proposed considering a single large indoor OWC IoT system with a large number of indoor OWC access points (APa).  In contrast, this work explores a novel two-tier \emph{multi-rate} SA design and develops a detailed system analysis and design for specific design assumptions at both OWC and RF tiers.

\section{Two-Tier Multi-Rate Slotted ALOHA}


We consider a set of $K$ indoor OWC IoT systems deployed across a wide area to connect indoor OWC IoT devices to their respective OWC APs within, e.g., a smart home environment. 
Each system contains $U$ IoT devices that access the OWC AP using SA protocol. Each IoT device transmits a fixed-length data packet during a slot of duration $T_{\rm  owc}$. During each slot, the OWC AP attempts to decode data packets from the received signal. 
We consider a receiver model for OWC AP where multiple packets originating at multiple active devices are decoded by exploiting the capture effect. The OWC APs operate as decode and forward (DF) relays, where the packets decoded during a single OWC slot are transmitted during the following RF slots as described below.

The $K$ indoor OWC IoT APs are connected to the BS via a LP WAN RF-based network. To be more specific, we assume the outdoor LP WAN is based on LoRa physical layer  and uses SA multiple access with slot duration $T_{ \rm rf}\leq T_{ \rm owc}$
\footnote{Slot synchronisation incurs slightly increased system complexity as compared to commonly used LoRaWAN which uses spread-ALOHA.}
LoRa PHY is based on a Chirp Spread Spectrum modulation scheme \cite{loraref}, improving robustness against interference and multipath fading.  
Besides, it uses quasi-orthogonal spreading factors (SFs),  meaning that the users with different SFs will  cause negligible inter-SF interference to each other \cite{loraref, R1}. 
If LoRa users are closer to the BS, transmission at lower SF will be adopted leading to the lower reception sensitivity, decreased processing gain, as well as shorter chirp symbol length and higher data rates (see Sec. II.2). Hence, based on distance $d_{\rm SF}$ between OWC IoT APs and the BS, the BS is in charge to assign SF $\in \lbrace 7,8,\cdots,12\rbrace$ while maintaining constant bandwidth (${\rm BW}$). This affects the chirp symbol duration as $T_s=2^{\rm SF}/{\rm BW}$, as well as data rate and slot duration $T_{ \rm rf}$ \cite{loraref}.

\begin{figure}[!t]
\centerline{\includegraphics[width=3.5in]{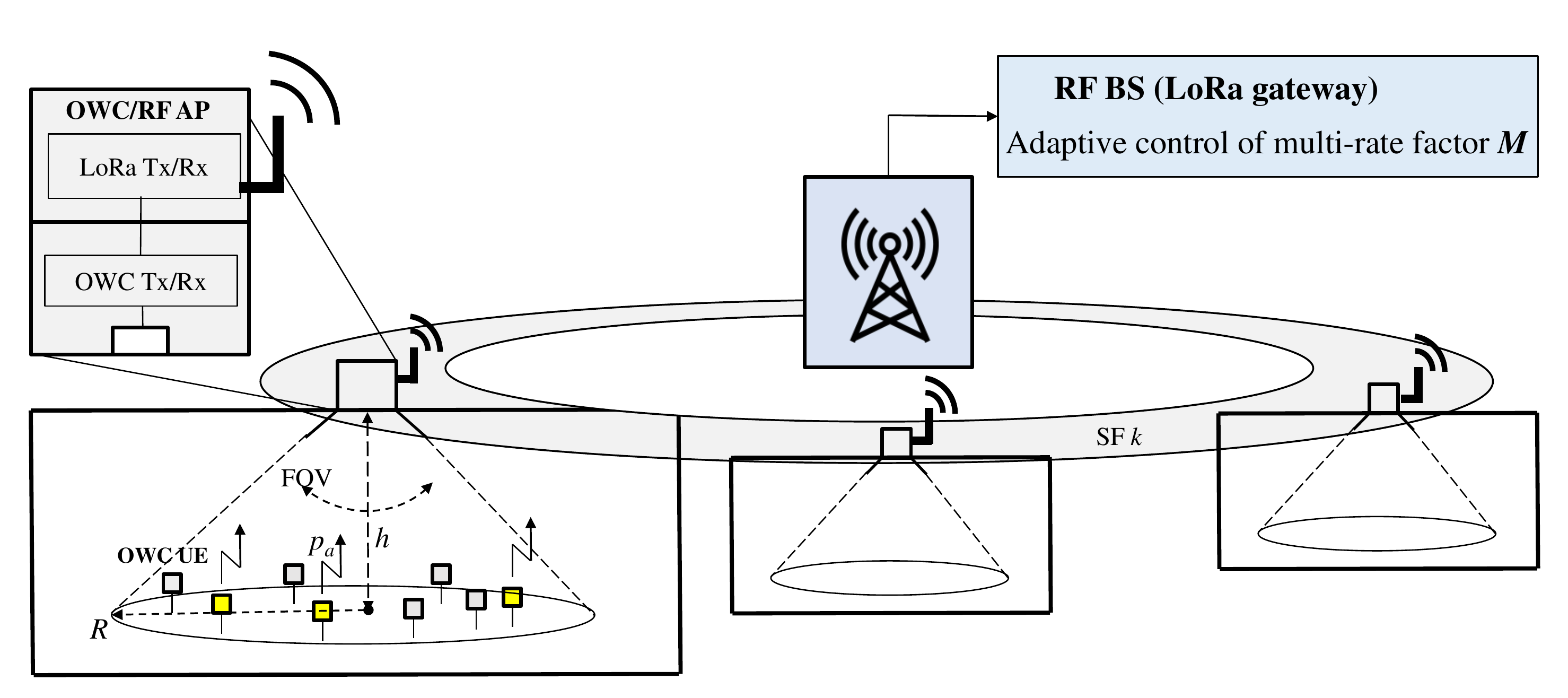}}
\caption{Two-Tier OWC/RF-Based IoT Network.}
\label{model}
\end{figure}

Since we assume negligible inter-SF interference, we will focus on a set of indoor OWC APs that use the same SF while transmitting to the LP WAN BS. This means that all $K$ indoor OWC APs are located at the approximate distance $d_{\rm SF}$ from the gateway (more precisely, being inside  the ring of the  radius $d_{\rm SF} \pm 1$ km), as it is depicted in Fig. \ref{model}. 
Additionally, we assume that the OWC slot duration $T_{ \rm owc}$ is dependent on the slot duration $T_{ \rm rf}$ at the LP WAN as $T_{ \rm owc}=M \cdot T_{\rm  rf}$ $(M \in \mathbb{N} /{0})$. For a given BW, the value of SF will determine the chirp symbol length $T_s$, and consequently the RF slot duration $T_{\rm rf}$, and data rate of RF transmission. 
By adjustment and adaptive control of multi-rate factor $M$, the LoRa BS effectively controls the slot rate as $T_{ \rm owc}=M \cdot T_{\rm  rf}$ in the OWC sub-systems, as well as a data rate of OWC transmissions. 
In this way, the LoRa network is capable to keep a constant BW while providing adaptive data rates determined by  $M$ in our overall system model.
 For simplicity, we assume the indoor OWC systems are slot-synchronized, and the outdoor RF system slot is aligned with $M$ consecutive OWC system slots. Clearly, the RF system operates at $M$ times higher data rate.

\subsubsection{Indoor OWC IoT Network}

We consider the SA scheme with Bernoulli arrivals  to model the distribution of the number of active users in a given slot for all indoor OWC  systems. This means that each of $U$ IoT devices  becomes activated in a given slot with  probability $p_a$ independently of any other user. Assuming that $U_a^{(i)}$ represents the number of activated users in the $i$-th OWC IoT system in a given slot ($i=1,\cdots,K$), it holds $ \mathbb{P} [ U_a^{(i)} = u ] = { U \choose u } p_a^u ( 1 - p_a)^{U-u}$
with  $U_a^{(i)}$ being a binomial random variable with mean $ U p_a$.

The IoT devices employ LED-based transmitters  operating in the IR spectrum and employing intensity modulation with on-off keying  to satisfy non-negative constraint. All OWC APs include photodetectors as OWC receivers based on direct detection of the light intensity.
In each of the $K$ indoor OWC  systems only the line-of-sight (LoS)  component of the optical signal is considered since  the reflected signals energy is significantly lower \cite{OWC_MATLAB}. A Lambertian channel model for the LOS link is adopted\footnote{Based on the Lambertian law for modeling the optical LoS link, the intensity of the optical signal between the $j$-th  IoT device  and the $i$-th  OWC AP ($i=1,\cdots,K$, $j=1,\cdots,U_a^{(i)}$) is defined as 
\begin{equation*}
 h_{i,j} =\!\! \left\{ {\begin{array}{*{20}{c}} \!\!\!\!\!
\frac{ A\left( m + 1\right) \mathcal{R} T_g g\left( \psi_{i,j} \right)}{2\pi d_{i,j}^2}\cos^m\left( \theta_{i,j} \right)\cos \left( \psi_{i,j} \right), 0 \leq \psi_{i,j} \leq \Psi\\
{0,}  \quad \quad \quad \quad \quad \quad \quad \quad \quad \quad\quad\quad {\rm otherwise} \\
\end{array}} \right.\!\!\!\!,
\label{I_ij}
\end{equation*}
where $\Psi $ is the field of view (FOV)  of the receiver, $A$ is the physical area of the detector, $\mathcal{R}$ is the responsivity, $T_g$ denotes the gain of the optical filter, while $g( \psi_{i,j})=l^2 /\sin^2\left( \Psi \right)$, for $0 \leq \psi_{i,j} \leq \Psi$, is the the optical concentrator with $l$ being  the refractive index of lens at a photodetector. The Lambertian order of the light source is  $m = -\ln 2/ \ln \left( \cos \Phi_{1/2} \right)$, where $\Phi_{1/2}$ represents the semi-angle at half power of LED. Distance, irradiance angle and  incidence angle between the $j$-th IoT user  and the $i$-th OWC AP are defined by $d_{i,j} , \theta_{i,j} $ and $\psi_{i,j}  $, respectively.
}, as well as the assumption that $U$  IoT devices are uniformly placed on a horizontal plane\footnote{For uniform distribution of the IoT users placed on horizontal plane  within the circle of radius $R$,  the PDF of the  radial distance $r$ of a randomly placed user from an OWC AP receiver is $f_{r}\left( r \right) = 2r/R^2,\quad 0\leq r \leq R$
.}  within the circle of radius $R$. The  SNR of the OWC link between the $j$-th user and the $i$-the OWC AP is  defined as $\gamma_{\rm owc}^{i,j} = P_t^2 h_{i,j}^2 \eta ^2 / \sigma_n^2$ $(i=1,\cdots,K$, $j=1,\cdots,U_a^{(i)})$\footnote{In the following, we omitted the indexes $i$ and $j$ for better clarity.}, where  $P_t$ is the transmitted optical power, $\eta$ is the  optical-to-electrical conversion
efficiency, while the Gaussian noise variance is determined as $\sigma_n^2=N_0 B$   with $N_0$ and $B$ being the noise spectral density  and the system  bandwidth, respectively. After utilization of the random variables  transformation techniques, the probability distribution function (PDF) of signal-to-noise ratio  (SNR)  $\gamma_{\rm owc}$ is derived as \cite{ref12,ref13}
\begin{equation}
f_{\gamma_{\rm owc}}\left( \gamma  \right) = \frac{\left( \mu _{\rm owc}\mathcal X ^2 \right)^{  \frac{1}{m + 3}}}{R^2\left( m + 3 \right)}\gamma^{ - \frac{m + 4}{m + 3}},\quad \gamma_{\min }\leq \gamma \leq \gamma_{\max },
\label{pdfVLC}
\end{equation}
where $L$ is the distance  between horizontal plane where users are located and the ceiling,  $\mathcal X = \frac{ A\left( m + 1 \right) \mathcal{R}}{2\pi}T_g g\left( \psi \right)L^{m + 1}$, $\gamma_{\min} = \frac{\mu_{\rm owc}  \mathcal X^2}{{\left( R^2 + L^2 \right)}^{m + 3}}$,   $\gamma_{\max} = \frac{\mu_{\rm owc} \mathcal X^2}{L^{2 \left(m + 3\right)}}$ and $\mu_{\rm owc}= P_t^2  \eta ^2 / \sigma_n^2$. 
The cumulative distribution function (CDF) of the  SNR is  \cite{ref13}
\begin{equation}
F_{\gamma_{\rm owc}}\!\!\left( \gamma  \right)\!\! =\!\! \left\{ {\begin{array}{*{20}{c}}
\!\!\!{1\! +\! \frac{L^2}{R^2}\! - \!\frac{1}{R^2}{{\left( {\frac{ \mu _{\rm owc}\mathcal X ^2  }{\gamma}} \right)}^{\frac{1}{{m + 3}}}},}  \gamma_{\min}\!\!\leq \!\!\gamma \leq\gamma_{\max}\\
{1,}  \quad \quad \quad \quad \quad \quad \quad \quad \quad \quad \gamma > \gamma_{\max}\\
\end{array}} \right.\!\!\!\!.
\label{cdfVLC}
\end{equation}

\begin{figure*}[!t]
\normalsize
\centering
 \includegraphics[width=15cm,height=7cm]{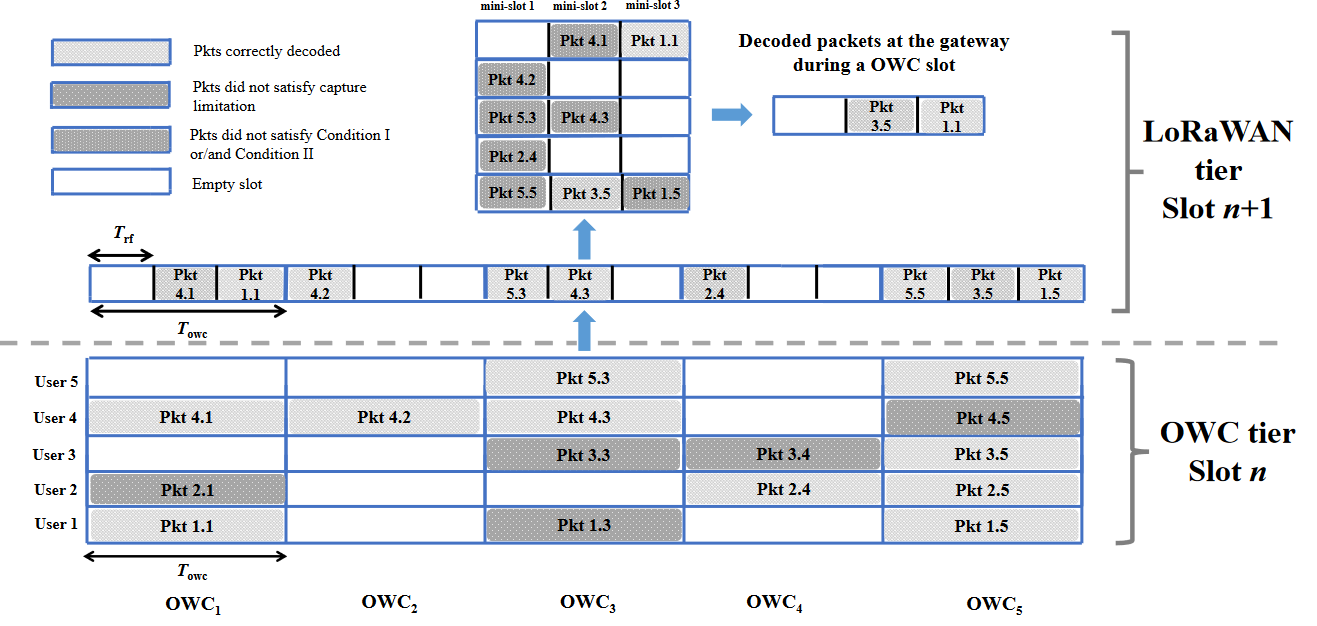}
\caption{Example of packet transmission in a OWC slot.}
\label{slots}

\vspace*{4pt}
\end{figure*}

\textbf{SA with Capture and MPR}: 
Regarding the  OWC tier, SA with capture with MPR is considered. 
For SA with capture, a packet can be decoded only if  the received SNR over OWC  link, $\gamma_{\rm owc}$ is above a  previously determined threshold, denoted by $\gamma_{\rm th}$. The probability that packet will not be  decoded  can be determined by the CDF defined in (\ref{cdfVLC}) as
  \begin{equation}
  { \rm P_{er}} =\mathbb{P}\left[ \gamma_{\rm owc} < \gamma_{\rm th}  \right]   = F_{\gamma_{\rm owc}}  \left( \gamma_{\rm th}  \right).
  \label{capture}
  \end{equation}
This kind of SA with capture within OWC can be easily implemented by exploiting different spread spectrum techniques with the aim to reduce the impact of the interference and to improve security of transmission  \cite{SS1}.

Additionally, MPR is also considered as multiple packets from  multiple active devices in the same OWC IoT system can be decoded by corresponding OWC AP. 
We set the maximal number of packet to be possibly decoded to be equal to $M$ -  the  ratio between slot durations in OWC and RF tiers, determining the maximal load of the LP WAN RF network. 

\subsubsection{Outdoor Wide-Area RF  IoT Network}

If the received power of the packet is among $M$  highest  of all received packets at corresponding OWC AP, and the capture condition is satisfied, the packet will be  correctly received and randomly assigned to one of $M$ newly formed slots. Also, only one packet can be assigned to each of the new slots.
All OWC AP, acting as DF relays, re-encode and forward data packets to the LoRa BS during assigned new slot. 
RF tier also employs SA approach with different slot duration $T_{\rm rf}$ as defined earlier.

 The number of active devices in LoRa system is now determined by the OWC IoT systems and the value of $M$. While keeping a constant BW, the gateway adjusts the SF based on the distance $d_{\rm SF}$. As it is presented in Table I, the SF has impact  on a bit-rate (for BW=125 kHz), as well as the time-on-air of a transmission and receiver sensitivity (determined by the SF specific threshold  $q_{\rm SF}$). As it was mentioned earlier, for  a given value of  BW=125 kHz, adjustment of SF will affect  the RF slot duration $T_{\rm rf}$ and data rate of RF transmission.
Furthermore, a LoRa BS is able to perform adaptive control of parameter $M$, which directly results in the design of the  slot rate $T_{\rm owc}$ in the  OWC sub-systems as $T_{\rm owc}=M \cdot T_{\rm rf}$, and consequently on   a data rate of OWC transmissions.





If a packet from OWC IoT system is successfully decoded, it is forwarded to final gateway with  transmit power $P_{\rm rf}$  over  Rayleigh distributed  block flat-fading channel. The instantaneous SNR of a single end-device can be defined as $\gamma_{\rm rf} = P_{\rm rf} g_{\rm SF} h_{\rm rf}^2 / \sigma_{\rm rf}^2$, where $h_{\rm rf}$ is the fading amplitude of the RF link modeled by Rayleigh distribution, $g_{\rm SF}=(\lambda / 4\pi d_{\rm SF})^n$ is path loss attenuation function determined  by the Friis transmission equation, where $\lambda$ is the carrier wavelength,  $n$ is the path loss exponent to be equal to 2.7 and 4 in urban and sub-urban environments, respectively, and the range $d_{\rm SF}$ is determined by SF in Table I. Additive white Gaussian noise variance $\sigma_{\rm rf}^2$ is determined as $\sigma_{\rm rf}^2= -174 + {\rm NF}+10{\rm log BW}$ dBm, where 
NF  is the receiver noise figure  \cite{loraref}.

At the BS, a desired packet will be  successfully decoded if outage does not happen.  In order to determine the outage probability of an uplink transmission, we adopt the
 model that assumes the outage will happen at the BS if either of following two conditions are satisfied \cite{loraref}:
 
\textbf{Condition I}:
The instantaneous SNR of the received packet  is below the SF specific threshold $q_{\rm SF}$ given in Table I, i.e., 
  \begin{equation}
{ \rm P_{out}^I}  =\mathbb{P}\left[{ \gamma_{\rm rf} < q_{\rm SF}}  \right] =\mathbb{P}\left[ \frac{P_{\rm rf} g_{\rm SF} h_{\rm rf}^2}{\sigma_{\rm rf}^2}  < q_{\rm SF}  \right] 
  \label{Condition}.
  \end{equation}
Since Rayleigh distributed block flat-fading channel is assumed for second tier, $h_{\rm rf}^2 $ represents the   channel
gain modelled by exponential distribution  with mean one. The first outage condition can be determined as
  \begin{equation}
{ \rm P_{out}^I} =\mathbb{P}\left[  h_{\rm rf}^2  < \frac{\sigma_{\rm rf}^2 q_{\rm SF} }{P_{\rm rf} g_{\rm SF}}  \right] = 1 - {\rm exp} \left( \frac{\sigma_{\rm rf}^2 q_{\rm SF} }{P_{\rm rf} g_{\rm SF}} \right) 
  \label{ConditionI}.
  \end{equation}

\begin{table}[t]
\centering
\caption{\bf LoRa characteristics \cite{loraref}}
\begin{tabular}{ccccc}
\hline
SF  &  Bit-rate &  $T_{\rm rf}$ &  $q_{\rm SF}$ & $d_{\rm SF}$ \\
\hline
7 & 5.47 kb/s & 36.6 ms & -6 dBm & 1 km ($\pm $1)  km  \\
\hline
8 & 3.13kb/s  & 64 ms & -9 dBm & 3 km ($\pm $1) km \\
\hline
9 & 1.76 kb/s& 113 ms & -12 dBm &  5 km ($ \pm $1) km \\
\hline
10 & 0.98 kb/s & 204 ms & -15 dBm & 7 km ($\pm$1) km \\
\hline
11 & 0.54kb/s  & 372 ms & -17.5 dBm & 9 km ($\pm $1) km \\
\hline
12 & 0.29kb/s  & 682 ms & -20 dBm & 11 km ($\pm $1) km  \\
\hline
\end{tabular}
  \label{table}
\end{table} 

\textbf{Condition II}:
 Assuming that the desired signal is the strongest one, by this condition, it should be at least $\epsilon$ times stronger than any other signal with the same SF:
  \begin{equation}
{ \rm P_{out}^{II}} =\mathbb{P}\left[  \frac{\gamma_{\rm rf} }{\gamma_{\rm rf}^* } \leq \epsilon  \right] 
  \label{ConditionII},
  \end{equation}
  where $\gamma_{\rm rf}^*$ represents the instantaneous SNR of interfering signal transmission of the same SF.
We adopt $\epsilon=4$, i.e., Condition II assumes that the desired signal is at least 4 times (6 dB) stronger than any other signal with the same SF.

A graphical example of proposed system transmission is presented in Fig.~ \ref{slots}, assuming $K=5$ OWC IoT sub-systems each containing $U=5$ users, and multi-rate factor $M=3$. Depending on the value of $p_a$, during the $n$-th slot, a subset of users in each OWC system will be active. In our example, in the first OWC system, Users 1, 2 and 4 will generate the data, in the second OWC system,  User 2 will be active, and so on (see Fig.~ \ref{slots}). Packets 2.1, 1.3, 3.3, 3.4 and 4.5 will be erased at the OWC AP since their   SNRs are assumed lower than $\gamma_{\rm th}$ due to the SA scheme with capture. Up to $M$ packets can be decoded at each OWC AP, which  are randomly assigned to $M$ slots at the LP WAN tier. When number of decoded packets exceeds $M=3$ ( e.g., the 5-th OWC system in our example), packets with higher received power have priority to be forwarded (packet 2.5 will be erased due to the lowest SNR). During the $(n+1)$-th OWC slot, the second tier uses $M$ RF slots to forward the packets decoded by each of $K$ OWC APs during the previous $n$-th OWC slot. After LP WAN transmission, some of the packets will be erased in all mini-slots due to Condition I and Condition II. For example, in Fig.~ \ref{slots}, packets 4.2, 4.1, and so on are erased, while packets 3.5 and 1.1 will be the only to be decoded.

\section{Numerical results and discussion}


In this section, we present numerical results obtained by Monte Carlo (MC) simulations. We consider throughput depending on activation probability $p_a$ of users in indoor OWC clusters. Note that the throughput is defined by averaging the number of decoded packets per RF slot. For this reason, the maximum throughput will not depend on \textit{M}. We note that the  throughput will linearly increase with M if it is defined as an average number of decoded packets per OWC slot.

 \textbf{Simulation setup:} Following values for the  parameters are adopted:  FOV = $90^{\circ}$, $A=1~{\rm cm}^2$,  $\mathcal R=0.4~{\rm A}/{\rm W}$,  $T_g =1$,$l =1.5$, $\eta=0.8$,  $N_0=10^{-21}~{\rm W}/{\rm Hz}$ and $B=200~{\rm kHz}$. Further, $\gamma_{\rm th}=0$ dB, $P_t = 10$  dBm, NF $=6$ dB,  BW $=125$ kHz, $n=2.7$, while  $P_{\rm rf}=14$ dBm as maximal transmit power  in Europe, and $\lambda=c/f_c$ when $f_c=868.1$ MHz.

\begin{figure}[!t]
\centerline{\includegraphics[width=3.5in]{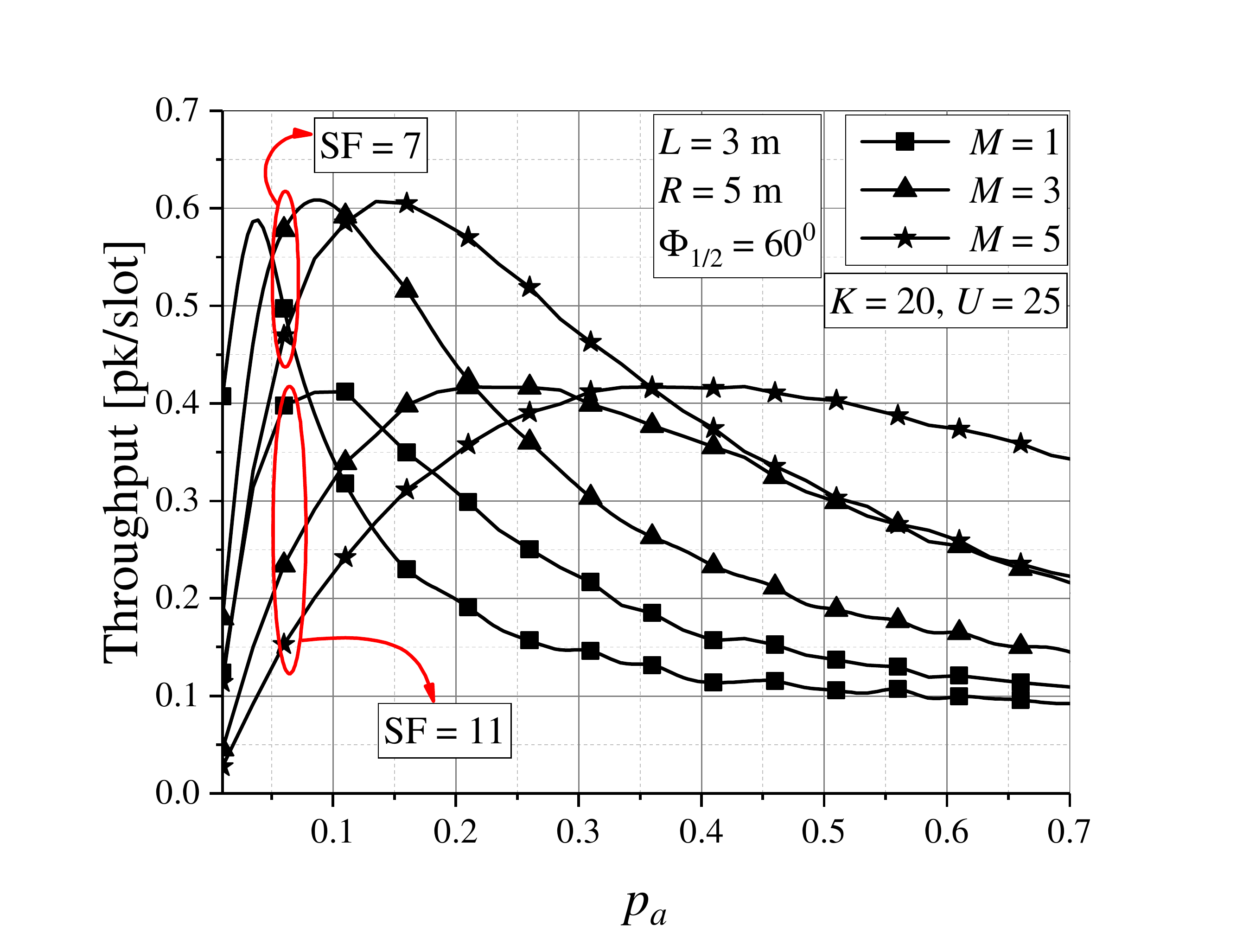}}
\caption{Throughput vs. $p_a$ for different $M$.}
\label{Fig_M}
\end{figure}

\textbf{Numerical results:}  Fig.~ \ref{Fig_M} presents throughput versus $p_a$ for different multi-rate parameter $M$ when the OWC IoT systems are very close to the gateway (SF $=7$) and when their distance is larger (SF $=11$). Recall that the distance $d_{\rm SF}$ determines both the spreading factor SF and the slot duration $T_{\rm rf}$ (Table I), while the multi-rate factor $M$ determines the slot duration $T_{\rm owc}$. In general, the values of SF and $M$ directly affect data rates of both tiers. From Fig. \ref{Fig_M} we can conclude that maximal throughput occurs for certain optimal value of $p_a$, which depends on both SF and parameter $M$.
Additionally, for larger $p_a$, which corresponds to increased number of active users in all OWC IoT systems, higher $M$ results in higher system throughput. As detailed later in the section, by providing adaptive data rates controlled by $M$, significant improvement of the system performance can be achieved with respect to the LoRa RF range and distances $d_{\rm SF}$.  

\begin{figure}[!t]
\centerline{\includegraphics[width=3.5in]{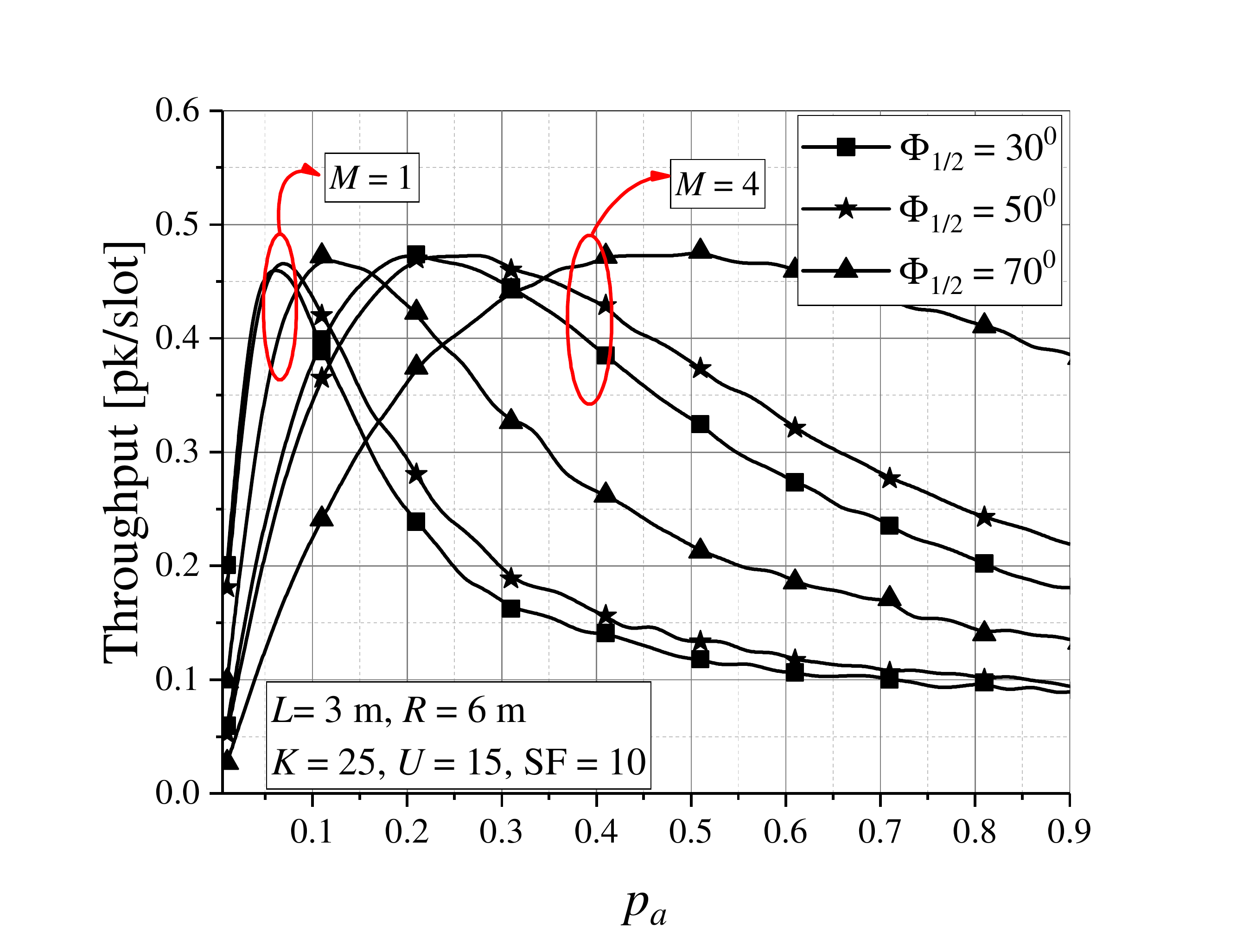}}
\caption{Throughput vs. $p_a$ for different values of semi-angle $\Phi_{1/2}$.}
\label{Fig_SA}
\end{figure}

\begin{figure}[!t]
\centerline{\includegraphics[width=3.5in]{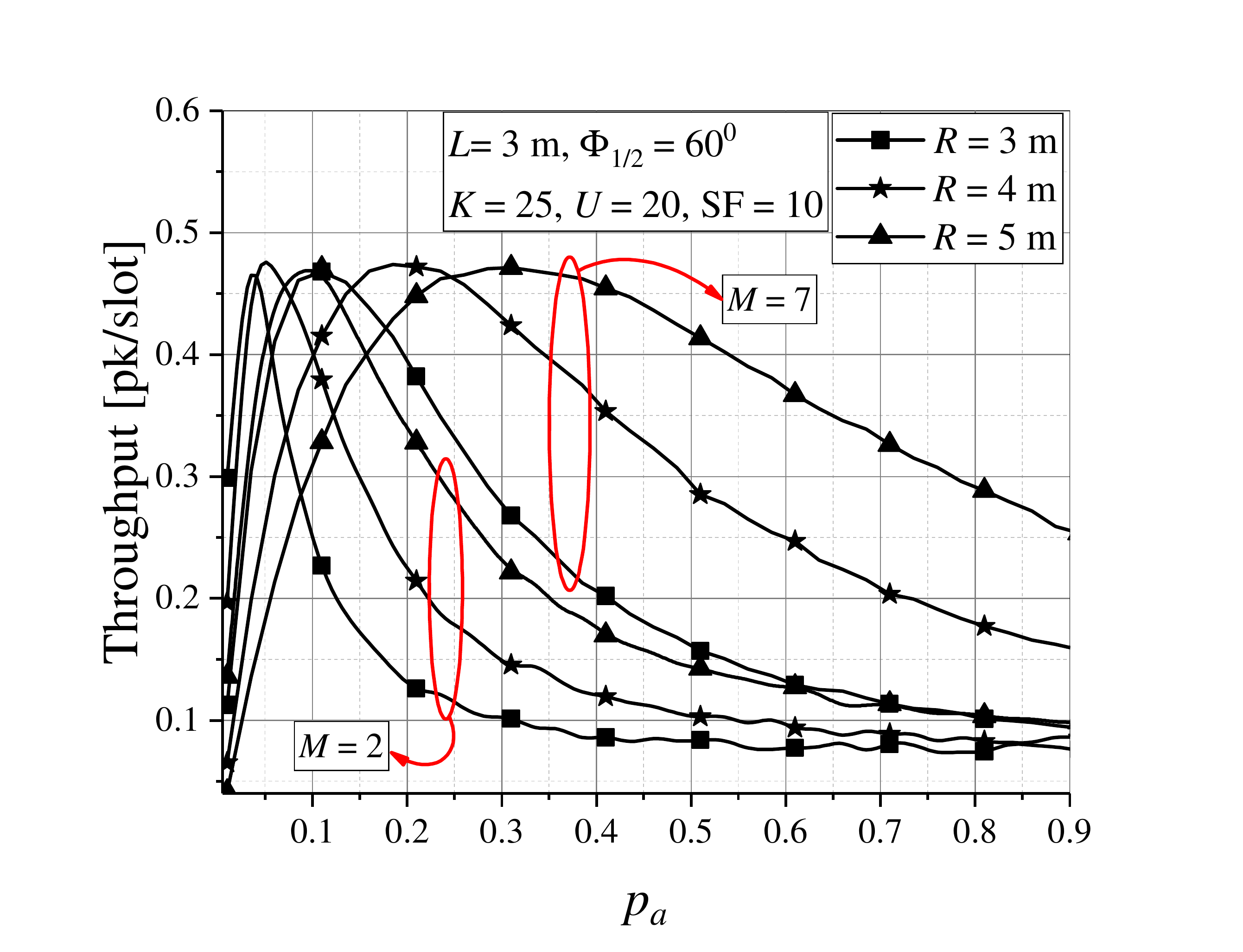}}
\caption{Throughput vs. $p_a$ for different values of radius $R$.}
\label{Fig_R}
\end{figure}

Further, in Figs.  \ref{Fig_SA}  and  \ref{Fig_R} we observe how indoor OWC  parameters  affect the    OWC/RF  system performance for different  data rates at the tiers determined by the parameter $M$.


Impact of the semi-angle $\Phi_{1/2}$ on the overall throughput is presented in Fig. \ref{Fig_SA} for different $M$. The value of  $\Phi_{1/2}$ determines how the light intensity is distributed over the receiving plane. When $\Phi_{1/2}$ is larger, LEDs output  beam  will be wider, thus the received power at the AP will be greater. In that case, the OWC IoT users far distant from OWC AP are likely to satisfy capture condition. With lowering $\Phi_{1/2}$, the OWC users will radiate narrower optical beams, and distant OWC users  will not satisfy capture condition. For that reason, when number of active users (i.e., $p_a$) is larger, the  throughput improves when $\Phi_{1/2}$ is lower, since capture limitation  goes in favor of reducing the number of decoded packets and achieving optimal number of successfully decoded ones. Overall, for different values of $\Phi_{1/2}$, the maximal throughput depends on the values of $p_a$ and  the parameter $M$.

Regarding the impact of  the radius $R$  on the overall throughput,   similar conclusions can be drawn from Fig. \ref{Fig_R}. 
The radius $R$ determines the size of the area where OWC IoT devices are located, which reflects in the distances between the devices and the OWC AP, and consequently on the received power at the AP. For example, when $R$ is smaller, users will be close to the  AP thus received packets will be decoded due to enough power to satisfy capture condition. Maximal throughput also exists for optimal value of $p_a$, which depends on   $R$  and $M$. 

\textbf{Adaptive control of multi-rate factor \textit{M}:}
The above throughput analysis indicates a clear trade-off between the OWC system parameters ($R$,$\Phi_{1/2}$), $p_a$, SF and $M$ for the system performance optimization. Given that most of the parameters may be beyond control of the system designer, we note that controlling $M$ is an elegant way for the proposed two-tier system to operate at favourable throughput.

From Fig. \ref{Fig_M}, we note that for low number of active users (e.g., $p_a<0.05$ when SF=7 and $p_a<0.1$ when SF=11), the throughput is maximised for $M=1$, i.e., when the data rates of both tiers are equal. If the number of active users increases, introducing the multi-rate functionality improves the system performance. For example, for $p_a<0.2$ and for short OWC IoT APs and BS distance (SF=7 and consequently $T_{\rm rf}=36.6$ ms), from Fig. \ref{Fig_M} we observe that LoRa BS 
should adjust multi-rate factor to $M=3$, i.e., to adjust the OWC slot rate to $T_{ \rm owc}=M \cdot T_{\rm  rf}=109.8$ ms, in order to maximise the system throughput. For a given OWC parameters, the LoRa BS could use a simple lookup table to select the value of $M$ and consequently the OWC slot rate  $T_{ \rm owc}$ based on employed SF and estimated activity probability $p_a$ (that can be obtained from OWC APs via user activity detection methods \cite{r19}).

Figs. \ref{Fig_SA} and \ref{Fig_R} present the effect of adaptive control of $M$ on the system throughput as a function of OWC system parameters. For a given OWC system parameters ($ \Phi_{1/2}$, $R$), the LoRa BS will select the value of $M$ that achieves the maximal throughput for a given $p_a$. This control mechanism can be implemented at LoRa BS as a simple lookup table that maps estimated $p_a$ to the optimal value of $M$.

\ifCLASSOPTIONcaptionsoff
  \newpage
\fi

\end{document}